\documentclass[aps,preprint]{revtex4}%
\usepackage{amsfonts}
\usepackage{amsmath}
\usepackage{amssymb}
\usepackage{graphicx}%
\begin{document}
\title{{\LARGE GENERAL RELATIVISTIC SINGULARITY-FREE COSMOLOGICAL MODEL }}
\author{Marcelo Samuel Berman$^{1}$}
\affiliation{$^{1}$Instituto Albert Einstein / Latinamerica - Av. Candido Hartmann, 575 -
\ \# 17}
\affiliation{80730-440 - Curitiba - PR - Brazil email: msberman@institutoalberteinstein.org}
\date{(Original: \textit{circa }2001; Last version: 25 February, 2009)}

\begin{abstract}
We "explain", using a Classical approach, how the Universe was created out of
\ "nothing" ,i.e., with no input of initial energy nor mass. The inflationary
phase, with exponential expansion, is accounted for, automatically , by our
equation of state for the very early Universe. This is a Universe with
no-initial infinite singularity of energy density.

\end{abstract}
\maketitle

\begin{center}

{\LARGE GENERAL RELATIVISTIC SINGULARITY-FREE COSMOLOGICAL MODEL }

DOI 10.1007/s10509-009-0024-8

Marcelo Samuel Berman
\end{center}

{\large \bigskip1 INTRODUCTION}

\bigskip Berman and Trevisan (2001), suggested that with a "correct" equation
of state, the Universe, as depicted with Robertson-Walker's metric, and the
Einstein's field equations, appeared to have been created out-of-nothing. In
the present paper, we derive an equation of state for the very early Universe,
which consequence is a singularity-free exponential big-bang, such that
inflation may be, in fact, the primordial phase of the Universe.

\bigskip

The science of cosmology (Weinberg, 1972; 2008) has progressed up to a point
where it is possible to make a valid picture of the first moment of the
Universe. It is widely accepted that the creation of the Universe should be
attributed to Quantum fluctuations of the vacuum. As a consistent Quantum
gravity theory is not yet available, cosmologists employ General Relativity,
or other classical theories, in order to study the evolution of the Universe
after Planck 's time \ ( \ $t$ $\cong10^{-43}s$ \ )\ \ .\ \ \ \ \ \ \ \ \ 

\bigskip

The present author will try to offer now a Classical picture covering the
creation of the Universe (Berman, 2007, 2007a, 2008). The only thing that we
need to admit, is that Einstein%
\'{}%
s field equations yield the average values for the quantities that, in the
Quantum Universe, when $t<10^{-43}s$ \ , fluctuate quantum-mechanically around
those average values, somehow like the Path Integral theory of Feynman (1965)
admits paths that fluctuate around the average trajectories given by the
Classical theory. Throughout this paper, when we refer to the \textit{alias}
Quantum Universe, all quantities should be understood as given by the most
probable values of those quantities, even if this is not explicitly stated.

Along a rather similar situation, Kiefer, Polarski and Starobinsky (1998) have
shown that although according to the inflationary scenario for the very early
Universe, all inhomogeneities in the Universe are of genuine quantum origin,
clearly no specific quantum mechanics properties are observed when looking at
this inhomogeneities and measuring them. It looks like we can get along
without a Quantum gravity theory, and still \textquotedblright explain
\textquotedblright\ the origin of the Universe.

It was shown by Cooperstock and Israelit (1995) that the total energy of the
Robertson-Walker's Universe is zero. The total energy is the sum of the
positive matter energy plus the negative energy of the gravitational field,
and it is zero. Thus, in the very creation moment there was no supply of
energy to the Universe, because it was not needed (see Feynman, 1962-63). We
may look at the subject like follows: in the very beginning there was nothing,
which we identify more or less philosophically, by assuming zero-total initial
energy. From conservation of energy, we may still consider that the Universe
is a zero-total energy entity.

\bigskip

\bigskip In Section 2, we show that the Robertson-Walker's metric is
zero-total energy valued. In Section 3, we introduce a solution for the very
early Universe, which is in fact an inflationary phase. In Section 4, we
discuss the singularity-free properties available with the present model.
Section 5 concludes the paper. \newline

\bigskip

\bigskip{\large 2 ZERO-TOTAL ENERGY OF THE UNIVERSE}

\bigskip\noindent Consider Minkowski's metric,

$ds^{2}=dt^{2}-\left[  dx^{2}+dy^{2}+dz^{2}\right]  $ \ \ \ \ \ \ \ \ \ \ \ \ \ \ \ \ \ \ \ \ \ \ \ \ \ \ \ \ \ \ \ \ \ \ \ \ \ \ \ \ \ \ \ \ \ \ \ \ \ \ \ \ \ \ (2.1)


This is an empty Universe, except for test particles. We agree that its total
energy is {zero} (Weinberg, 1972).


Now consider the expanding flat metric:

$ds^{2}=dt^{2}-R^{2}(t)\left[  dx^{2}+dy^{2}+dz^{2}\right]  $\ \ \ \ \ \ \ \ \ \ \ \ \ \ \ \ \ \ \ \ \ \ \ \ \ \ \ \ \ \ \ \ \ \ \ \ \ \ \ \ \ \ \ \ \ \ (2.2)


Here, $R(t)$ is the scale-factor. At any particular instant of time, $t=t_{0}
$ , we may define new variables, by the reparametrization,

$dx^{\prime}{}^{2}\equiv R^{2}\left(  t_{0}\right)  dx^{2}$\ \ \ \ \ \ \ \ \ \ \ \ \ \ \ \ \ \ \ \ \ \ \ \ \ \ \ \ \ \ \ \ \ \ \ \ \ \ \ \ \ \ \ \ \ \ \ \ \ \ \ \ \ \ \ \ \ \ \ \ \ \ \ \ \ \ \ \ \ \ \ \ \ (2.3)

$dy^{\prime}{}^{2}\equiv R^{2}\left(  t_{0}\right)  dy^{2}$\ \ \ \ \ \ \ \ \ \ \ \ \ \ \ \ \ \ \ \ \ \ \ \ \ \ \ \ \ \ \ \ \ \ \ \ \ \ \ \ \ \ \ \ \ \ \ \ \ \ \ \ \ \ \ \ \ \ \ \ \ \ \ \ \ \ \ \ \ \ \ \ \ (2.4)

$dz^{\prime}{}^{2}\equiv R^{2}\left(  t_{0}\right)  dz^{2}$\ \ \ \ \ \ \ \ \ \ \ \ \ \ \ \ \ \ \ \ \ \ \ \ \ \ \ \ \ \ \ \ \ \ \ \ \ \ \ \ \ \ \ \ \ \ \ \ \ \ \ \ \ \ \ \ \ \ \ \ \ \ \ \ \ \ \ \ \ \ \ \ \ \ (2.5)\ \ \ \ \ \ \ \ \ \ \ \ \ \ \ \ \ \ \ \ \ \ \ \ \ \ \ \ \ \ \ \ \ \ \ \ \ \ \ \ \ \ \ \ \ \ \ \ \ \ 


$dt$'$^{2}\equiv dt^{2}$

\begin{center}

\end{center}


Then,


$ds^{2}=dt^{\prime2}-\left[  dx^{\prime2}+dy^{\prime2}+dz^{\prime2}\right]  $\ \ \ \ \ \ \ \ \ \ \ \ \ \ \ \ \ \ \ \ \ \ \ \ \ \ \ \ \ \ \ \ \ \ \ \ \ \ \ \ \ \ \ \ \ \ \ \ \ \ \ \ \ \ \ \ \ \ \ \ \ \ \ \ \ \ \ \ \ (2.6)


The energy of this Universe is the same as Minkowski's one, namely, \ \ $E=0$.
We remember that in energy calculations, the instant of time is fixed.

\bigskip\bigskip

\noindent\bigskip\noindent Consider now the metric:

$ds^{2}=dt^{2}-\frac{R^{2}(t)}{\left[  1+\frac{kr^{2}}{4}\right]  ^{2}}\left[
dx^{2}+dy^{2}+dz^{2}\right]  $\ \ \ \ \ \ \ \ \ \ \ \ \ \ \ \ \ \ \ \ \ \ \ \ \ \ \ \ \ \ \ \ \ \ \ \ \ \ \ \ \ \ \ \ \ (2.7)


Here, \ \ $k=0$ \ \ \ yields the flat case, already studied. When \ \ $k$ =
$\pm1$, \ \ \ we have finite closed or infinite open Universes.


We want to calculate its energy. We are allowed to choose the way into making
the calculation, so we choose a fixed value $\bar{r}$ of the radial
coordinate, for which we reparametrize the metric:\bigskip

\bigskip
$dx^{\prime i}{}^{2}\equiv\frac{R^{2}\left(  t_{0}\right)  dx^{i2}}{\left[
1+k\frac{\bar{r}^{2}}{4}\right]  ^{2}}$\ \ \ \ \ \ \ \ \ \ \ \ \ \ \ \ \ \ \ \ \ \ \ \ \ \ \ \ \ \ \ \ \ \ \ \ \ \ \ \ \ \ \ \ \ \ \ \ \ \ \ \ \ \ \ \ \ \ \ \ \ \ \ \ \ \ \ \ \ \ \ \ \ (2.8)

\ 

\ \ \ \ \ \ \ \ \ \ \ \ \ \ \ \ \ \ \ \ \ \ \ \ \ ( $i=1,2,3$\ )

\bigskip

For this value of \ \ $r=\bar{r}$, the reparametrized metric has zero energy
value, by the same token as above. Now we sum for all other values of \ $r$\ ,
obtaining an infinite sum of zeros, which yields a total energy of zero value.\ \ 

\bigskip

\noindent

{\large 3 GRACEFUL ENTRANCE INTO INFLATION}

We suppose that the Universe obeyed Einstein's field equations, and
Robertson-Walker's metric. As the field equations yield an expansion, the most
probable value of $R$ (the scale factor) increases, beginning with zero value.

\bigskip

Let us consider Robertson-Walker's metric:

\bigskip\hfill$ds^{2}=dt^{2}$ $-$ $\frac{R^{2}}{\left[  1+kr^{2}/4\right]
^{2}}[dr^{2}+r^{2}d\theta^{2}+r^{2}\sin^{2}\theta d\phi^{2}]$ $,\ $\hfill(3.1)

\bigskip\noindent where $k$ represents the tricurvature ( $0,+1,-1$).

We can check that, Einstein's field equations read, for the above metric, and
a perfect fluid energy-tensor:

\bigskip\hfill$\kappa\rho=3H^{2}+3kR^{-2}+\Lambda$ ,\hfill(3.2)

\bigskip\hfill$\kappa p=-2R^{-1}\ddot{R}-H^{2}-kR^{-2}-\Lambda$ \ ,\hfill(3.3)

\bigskip\noindent where $\rho,\kappa,\Lambda,p,and$ $H$, stand respectively
for the energy density, Einstein's gravitational constant, cosmological
constant, cosmic pressure, and Hubble's parameter $(H=\dot{R}/R)$ .

On solving (3.2) and (3.3), we employ Berman and Trevisan's solution (Berman
and Trevisan, 2001),

\bigskip\hfill$R(t)=exp($ $\beta t$ $)-exp(-\beta t)$ , \hfill(3.4)

\bigskip\noindent where \ \ $\beta$\ \ is a constant to be determined. The
reason for introducing such metric will be seen later; it obeys the following
good conditions:

\bigskip

(a) $\lim\limits_{t\rightarrow0}$ $R(t)=0$ \ \ \ ;

and,

\bigskip

(b) $\lim\limits_{t\text{ }>>0}$ $R(t)=$ \ exponential scale-factor
(inflation) \ \ \ .

\bigskip

We\ remember that Berman and Trevisan obtained (3.4), by imposing an equation
of state of the type,

\bigskip

$\qquad\qquad\qquad\qquad\qquad\qquad\qquad p=-\frac{1}{3}\rho$
\ \ \ \ \ \ \ \ \ \ \ \ \ \ , \ \ \ \ \ \ \ \ \ \ \ \ \ \ \ \ \ \ \ \ \ \ \ \ \ \ \ \ \ \ \ \ \ \ \ \ \ \ \ \ \ \ (3.5)

\bigskip

\bigskip but here, we leave the equation of state non-determined, for the time-being.

\bigskip

From the given solution, we obtain now, the expression for Hubble's parameter,

\bigskip

$\qquad\qquad\qquad\qquad\qquad H\equiv\frac{\dot{R}}{R}=\beta\left[
\frac{1+\text{ }e^{-2\beta t}}{1-\text{ }e^{-2\beta t}}\right]  $
\ \ \ \ \ \ \ \ \ \ \ \ \ \ \ \ \ \ \ . \ \ \ \ \ \ \ \ \ \ \ \ \ \ \ \ \ \ \ \ \ \ \ \ \ \ \ \ \ \ \ (3.6)

\bigskip

If necessary, we may also remember that,

\bigskip

$\qquad\qquad\qquad\qquad\qquad\qquad\ddot{R}=\beta^{2}R$
\ \ \ \ \ \ \ \ \ \ \ \ \ \ \ \ \ \ \ \ \ \ \ . \ \ \ \ \ \ \ \ \ \ \ \ \ \ \ \ \ \ \ \ \ \ \ \ \ \ \ \ \ \ \ \ \ \ \ \ \ \ \ \ (3.7)

\bigskip

We consider now the case \ $k=0$\ \ , for a flat Universe. The field equations become,

\bigskip

\hfill$\kappa\rho=3\beta^{2}\left[  \frac{1+\text{ }e^{-2\beta t}}{1-\text{
}e^{-2\beta t}}\right]  ^{2}+\Lambda$ \ \ \ \ \ ,\hfill(3.8)

\bigskip\hfill$\kappa p=-2\beta^{2}-\beta^{2}\left[  \frac{1+\text{
}e^{-2\beta t}}{1-\text{ }e^{-2\beta t}}\right]  ^{2}-\Lambda$
\ \ \ \ \ \ \ .\hfill(3.9)

\bigskip

We now can find the necessary equation of state: we sum (3.8) and (3.9)\ and obtain,

\bigskip

\hfill$p=-\rho+\frac{2\beta^{2}}{\kappa}\left\{  \left[  \frac{1+\text{
}e^{-2\beta t}}{1-\text{ }e^{-2\beta t}}\right]  ^{2}-1\right\}  $
\ \ \ \ \ \ \ .\hfill(3.10)\ \ 

\bigskip

\bigskip We may also write,

\bigskip

$\ \ \ \ \ \ \ \ \ \ \ \ \ \ \ \ \ \ \ \ \ \ \ \ \ \ \ \ \ \ \ \ \ \ \ \ p=-\frac
{1}{3}\rho-\frac{2}{\kappa}\left(  \beta^{2}+\frac{1}{3}\Lambda\right)  $
\ \ \ \ \ \ \ \ \ \ \ . \ \ \ \ \ \ \ \ \ \ \ \ \ \ \ \ \ \ \ \ \ \ \ \ (3.11)

\bigskip

We have possibly negative cosmic pressures, but always larger than $\ \ -\rho
$\ \ \ \ and smaller than \ \ $-\frac{1}{3}\rho$\ \ \ \ \ .

\bigskip

\bigskip For larger values of \ $t$\ \ , as we are in face of rapid expansion,
the negative exponential can be neglected from our solution for the
scale-factor. This solution shows a graceful entrance into inflationary epoch
(Guth, 1981).

We shall see bellow, that, the resulting solution that makes \ $R(0)=0$ \ \ at
\ \ $t=0$, which does avoid the problems associated with infinite energy
densities at the initial instant of the Universe, results in that Einstein's
field equations, may apply even when Quantum phenomena come to play a vital r\^{o}le.

Berman(1991; 1991a) worked the full field equations for lambda variable, for a
general perfect gas equation of state, and a perfect fluid energy-tensor.

Einstein%
\'{}%
s theory need not be substituted by Quantum Gravity theories in dealing with
Planck%
\'{}%
s Universe, because we are taking the care of the understanding that the
Classical scale-factor, is representing an average value, over the uncertain
quantum paths.

\bigskip

We have, so far, derived the equation of state that might apply for the very
early Universe, and have shown that it leads towards a graceful entrance into
inflationary epoch. For large \ $t$\ \ , we obtain from (3.8), (3.9) and
(3.10) the following limiting equation of state,

\bigskip

$\qquad\qquad\qquad\qquad\qquad p\cong-\rho=-\left[  3\beta^{2}+\Lambda
\right]  $ \ \ \ \ \ \ \ \ \ \ \ \ \ \ \ . \ \ \ \ \ \ \ \ \ \ \ \ \ \ \ \ \ \ \ \ \ \ \ \ \ \ \ \ \ \ \ \ (3.12)

\bigskip

\bigskip This is the usual inflationary result.

\bigskip

{\Large \bigskip4 SINGULARITY-FREE UNIVERSE}

\bigskip

\bigskip Now, we can check the creation instant, where \ $t\rightarrow0$\ .
Calling \ $M$\ \ the mass of the Universe, we have, for a tri-dimensional
volume \ $V$\ \ , that,\ 

\bigskip

$\qquad\qquad\qquad\qquad\qquad\qquad\lim\limits_{t\rightarrow0}$
$R(t)=0$\ \ \ \ \ \ \ \ \ , \ \ \ \ \ \ \ \ \ \ \ \ \ \ \ \ \ \ \ \ \ \ \ \ \ \ \ \ \ \ \ \ \ \ \ \ \ \ \ \ \ \ \ \ \ \ \ \ \ (4.1)

\bigskip

and,

\bigskip

$\qquad\qquad\lim\limits_{t\rightarrow0}$ $M=\lim\limits_{t\rightarrow0}$
$\rho V=\frac{4}{3}\pi$ $\lim\limits_{t\rightarrow0}$ $\rho R^{3}%
=\lim\limits_{t\rightarrow0}$ $\left(  1-e^{-2\beta t}\right)  ^{2}%
=0$\ \ \ \ \ \ \ \ \ \ \ \ . \ \ \ \qquad(4.2)

We find that the scale factor increases from its zero value when \ $t=0$
\ \ and \ \ $M=0$ \ \ while the total energy input was also zero. (Feynman,
1962-63) This is creation out of nothing. Nevertheless, we presuppose that
Einstein's equations are valid along with the Robertson-Walker's metric for
the Universe since \ $t=0$ onwards. In the Quantum Universe, of course,
Planck's constant \ $h$ \ would have a specific r\^{o}le.

It must be remarked that a non-null lambda constant value, and, then, also of
dark energy, in order to get a negative value for the cosmic pressure, points
to an accelerating present Universe. The origin of the lambda concept should
belong to the Quantum Gravity chapter of particle physics in the very early
Universe but Berman (2008) has hinted that the cosmological term may have a
Classical origin (as a centrifugal acceleration in a rotating evolutionary
Universe -- see Berman, 2008a).

\bigskip

We have seen that the Universe begins from a state of no-mass and no-energy.
But we still need to think about the energy density in the origin of time. I
suggest that we employ a time-varying cosmological "constant" term, given by,

\bigskip

$\ \ \ \ \ \ \ \ \ \ \ \ \ \ \ \ \ \ \ \ \ \ \ \ \ \ \ \ \ \ \ \ \ \ \ \Lambda
=\Lambda_{0}-3H^{2}f(t)$ \ \ \ \ \ \ \ \ \ \ , \ \ \ \ \ \ \ \ \ \ \ \ \ \ \ \ \ \ \ \ \ \ \ \ \ \ \ \ \ \ \ \ \ \ \ \ \ \ \ \ \ (4.3)

\bigskip

\bigskip where \ $\Lambda_{0}$\ \ is a constant, \ $H$\ \ is given by (3.6) as
it should be, and the new function \ \ $f(t)$\ \ must obey the condition,

\bigskip

$\ \ \ \ \ \ \ \ \ \ \ \ \ \ \ \ \ \ \ \ \ \ \ \ \ \ \ \ \ \ \ \ \ \ \ \lim
\limits_{t\rightarrow0}$ $f(t)=1$ \ \ \ \ \ \ \ \ \ \ . \ \ \ \ \ \ \ \ \ \ \ \ \ \ \ \ \ \ \ \ \ \ \ \ \ \ \ \ \ \ \ \ \ \ \ \ \ \ \ \ \ \ \ \ \ \ \ \ \ \ \ (4.4)

\bigskip

In this case, we shall have a finite energy density in the origin of time,
given by \ \ $\frac{\Lambda_{0}}{\kappa}$\ \ . Now we have removed the initial
singularity, while keeping intact the rest of the previous model. The analysis
of varying \ $\Lambda$\ \ , has been authoritatively clarified in a paper by
Overduin and Cooperstock (1998).

\bigskip

{\Large 5 CONCLUSIONS}

\bigskip

The above original model, seems not to have been considered by any author. We
obtained, from Robertson-Walker's metric, a scale-factor that begins from
zero-value at the initial time, while we showed that the initial energy and
mass are equally zero. The energy density at the origin has a fixed finite
value, if we introduce a convenient time-varying cosmological term. The exact
form of the function \ $f(t)$\ \ , is left for future analyses, except for the
condition \ (4.4).

\bigskip

As Raychaudhuri et al. (1992) have commented, \ the singularity problem is
currently considered the most outstanding problem of Theoretical Cosmology.
The usual Friedman expanding Universe was endowed with the \ \ \ $R=0$%
\ \ \ \ problem, at a finite time in the past. Richard Tolman and Arthur
Eddington had thought that the singularity was a mere consequence of an
oversimplification, i.e., the isotropy and homogeneity assumptions, but that
in more realistic models, it would disappear. The use of Raychaudhuri's
equation (Raychaudhuri, 1955; 1979) has clarified the establishment of
singularity theorems. The infinite values of physical variables at some
points, may be over-thrown if we "cut-out" from the model the relevant portion
of spacetime where the variables blow-up, and the truncated region may be
presented as a "regular" spacetime. But, of course, we found a better solution
in this paper.

\bigskip

Our present model is both appropriate and convenient.

\bigskip

\bigskip\bigskip{\Large Acknowledgements}

I am indebted to an anonymous referee of a previous manuscript, which made me
aware of wrongdoings in the original manuscript, which here are corrected.

\bigskip I also thank my intellectual mentor Prof. Fernando de Mello Gomide,
and, Nelson Suga, Marcelo F. Guimar\~{a}es, Antonio F. da F. Teixeira, Mauro
Tonasse, and I am also grateful for the encouragement by Albert, Paula, and
Geni. I offer this paper \textit{in memoriam of \ \ }M. M. Som

\bigskip

{\Large References}

\bigskip

Berman, M.S (1991) -- \textit{Cosmological Models with Variable Cosmological
Term}, Physical Review \textbf{D43}, 1075-8.

Berman, M.S (1991a) -- \textit{Cosmological Models with Variable Gravitational
and Cosmological "Constants"}. General Relat. and Grav., \textbf{23}, 465-9.

Berman, M.S (1994) -- Astrophysics and Space Science, \textbf{222, }235.

Berman, M.S (2006) -- \textit{Energy of Black-Holes and Hawking's Universe.
}In \textit{Trends in Black-Hole Research, }Chapter 5\textit{.} Edited by Paul
Kreitler, Nova Science, New York.

\begin{description}
\item Berman, M.S. (2006a) - \textit{Energy, Brief History of Black-Holes, and
Hawking's Universe. }In \textit{New Developments in Black-Hole Research},
Chapter 5\textit{.} Edited by Paul Kreitler, Nova Science, New York.
\end{description}

\bigskip Berman,M.S. (2007) - \textit{Introduction to General Relativity and
the Cosmological Constant Problem}, Nova Science, New York.

Berman,M.S. (2007a) - \textit{Introduction to General Relativistic and
Scalar-Tensor Cosmologies}, Nova Science, New York.

\bigskip Berman,M.S. (2008) - \textit{A Primer in Black-Holes, Mach's
Principle, and Gravitational Energy }, Nova Science, New York.

Berman,M.S. (2008a) - \textit{General Relativistic Machian Universe,
}Astrophysics and Space Science, 318, 273-277.

Berman,M.S; Marinho Jr, R.M. (2001) -- \textit{Astroph. Space Science}
{\textbf{278}}, 367.

Berman, M.S; Som,M.M. (1989) -- Phys. Letters, \textbf{142A,} 338.

Berman, M.S.; Trevisan, L.A. (2001) - \ \textit{On the Creation of the
Universe out of "nothing"}\ , Los Alamos Archives http://arxiv.org/abs/gr-qc/0104060

Cooperstock, F.I; Israelit, M. (1995) -- Found. of Physics, 25:(4), 631.

Feynman, R.P.(1962-63) -- \textit{Lectures on Gravitation}, \ class notes
taken by F.B.Morinigo; W.C. Wagner, Addison Wesley, Reading.

Feynman, R.P; Hibbs, A.R. (1965) -- \textit{Quantum Mechanics and Path
Integrals}, Mc-Graw-Hill, New York.

Guth,A. (1981) --Phys Rev D \textbf{23}, 247.

Kiefer,C.; Polarsky, D.; Starobinsky,A.A (1998) -- gr-qc/9802003.

Kolb, E.W.; Turner, M.S. (1990) - \textit{The Early Universe}, Addison-Wesley, N.Y.

Overduin, J.M.; Cooperstock, F.I. (1998) - Physical Review, \textbf{D58,} 043506.

Raychaudhuri, A.K. (1955) - \textit{Physical Review }\textbf{98}, 1123.

Raychaudhuri, A.K. (1979) - \textit{Theoretical Cosmology, }Oxford University
Press, Oxford.

Raychaudhuri, A.K., Banerji, S.; Banerjee, A. (1992) - General Relativity,
Astrophysics and Cosmology, Springer. New York.

Weinberg, S. (1972) - \textit{Gravitation and Cosmology, }Wiley, New York.

Weinberg, S. (2008) - \textit{Cosmology,} Oxford University Press, Oxford.

\end{document}